# High-performance thermochromic multilayer coatings with W-doped $VO_2$ nanoparticles dispersed in $SiO_2$ matrix prepared on glass at a low temperature

Jaroslav Vlček,[a,*] Michal Kaufman,[a] Elnaz M. Nia,[a] Jiří Houška,[a] Jiechao Jiang,[b] Radomír Čerstvý,[a] Stanislav Haviar,[a] Efstathios I. Meletis [b]

E-mail address: vlcek@kfy.zcu.cz (J. Vlček)

[a] Department of Physics and NTIS – European Centre of Excellence, University of West Bohemia, Univerzitní 8, 30100 Plzeň, Czech Republic

[b] Department of Materials Science and Engineering, The University of Texas at Arlington, Arlington, 76019, TX, USA

**ABSTRACT**: We report a high-performance thermochromic $VO_2$-based coating prepared by using a three-step process, consisting of magnetron sputter depositions of $SiO_2$ films and V-W films and their postannealing, on standard glass at a low substrate temperature of 350 °C without opening the vacuum chamber to atmosphere. It is formed by four layers of W-doped $VO_2$ nanoparticles dispersed in $SiO_2$ matrix. The coating exhibits a transition temperature of 33 °C with an integral luminous transmittance of 65.4% (low-temperature state) and 60.1% (high-temperature state), and a modulation of the solar energy transmittance of 15.3%. Such a combination of properties, together with the low temperature during preparation, fulfill the requirements for large-scale implementation on building glass and have not been reported yet.

## 1. Introduction

Vanadium dioxide ($VO_2$) exhibits a reversible phase transition from a low-temperature monoclinic $VO_2$(M1) semiconducting phase to a high-temperature tetragonal $VO_2$(R) metallic phase at a transition temperature ($T_{tr}$) of approximately 68 °C for the bulk material [1]. $T_{tr}$ can be lowered using doping of $VO_2$ with other elements (such as W [2.3]). The automatic response to temperature and the abrupt decrease of infrared transmittance with almost unchanged luminous transmittance at the transition into the metallic state make $VO_2$-based coatings a promising candidate for thermochromic (TC) smart windows reducing the energy consumption of buildings.

To meet the requirements for large-scale implementation on building glass (glass panes, or flexible glass and polymer foils laminated to glass panes), $VO_2$-based coatings should satisfy the following strict criteria simultaneously (see the review [3] and the works cited therein): an integral luminous transmittance $T_{lum} > 60\%$, a modulation of the solar energy transmittance $\Delta T_{sol} > 10\%$, $T_{tr}$ near room temperature, long-term environmental stability, and a more appealing color than the usual yellowish or brownish colors in transmission. Moreover, a maximum substrate temperature ($T_s$) during preparation (deposition and possible postannealing) should be near 400 °C or lower as required by industrially important substrates, such as soda-lime glass (SLG) or ultrathin flexible glass. Here, a major challenge is to achieve the high $T_{lum}$ and $\Delta T_{sol}$ at the aforementioned decreased values of $T_{tr}$ and $T_s$ [3-6]. To the best of our knowledge, the simultaneous fulfillment of these requirements has been reported for only two TC coatings up to now: a three-layer YSZ/V$_{0.855}$W$_{0.018}$Sr$_{0.127}$O$_2$/SiO$_2$ coating, where YSZ denotes the yttria-stabilized zirconia, with the average $T_{lum}$ = 62.2%, $\Delta T_{sol}$ = 11.2% and $T_{tr}$ =22 °C, which was prepared on glass at $T_s$ = 320 °C [5], and a composite W-VO$_2$@AA/PVP film, where AA and PVP denote the L-ascorbic acid and the polyvinylpyrrolidone, respectively, with the average $T_{lum}$ = 69.3%, $\Delta T_{sol}$ = 10.2% and $T_{tr}$ =34.5 °C, which was prepared on glass at $T_s$ = 60 °C [6].

Recently, valuable results in the area of TC VO$_2$-based films for future smart-window applications have been achieved [7-10]. A two-step process consisting of magnetron sputter deposition and postannealing in an oxidizing environment made it possible to produce discontinuous VO$_2$ structure, i.e., dispersed subwavelength VO$_2$ nanoparticles, on quartz and sapphire substrates. The layers of the subwavelength (to minimize light scattering) VO$_2$ nanoparticles with a lower refractive index ($n$) and extinction coefficient ($k$) in visible range than those of the compact VO$_2$ layer [7] exhibited very high $T_{lum}$ due to a decreased visible-range reflectance and absorption [7,10]. At the same time, the localized surface plasmon resonance (LSPR) on the surface of VO$_2$(R) metallic nanoparticles enhanced $\Delta T_{sol}$ significantly due to an increased absorption in the near-infrared region. However, the $T_{tr}$ values of these TC nanostructured materials are too high (53-66 °C), as the fabricated VO$_2$ nanoparticles have not been doped, and the $T_s$ values during the preparation of these materials are too high (450-600 °C) for industrially important substrates.

In this study, we present a high-performance (the average $T_{lum}$ = 62.8% and $\Delta T_{sol}$ = 15.3%) TC VO$_2$-based coating with a decreased $T_{tr}$ = 33 °C and enhanced environmental protection, which was prepared on conventional SLG at a low $T_s$ = 350 °C. It is formed by four layers of W-doped VO$_2$ nanoparticles dispersed in a SiO$_2$ matrix. We present and explain the design and phase composition of the coating, microstructure, elemental composition and surface morphology of the individual nanocomposite layers, the TC properties of the multilayer coatings with one to four nanocomposite layers and their synthesis performed without opening the vacuum chamber to atmosphere.

## 2. Experimental details

2.1. Coating preparation

Prior to application of the TC coating, the 1 mm thick SLG substrate was covered without any external heating by a 100 nm thick SiO$_2$ layer blocking Na diffusion from the glass (see the analysis of the Na diffusion from SLG [11]). Then, a 10 nm thick V-W film was deposited also without any external heating. Subsequently, the coating was annealed up to 350 °C for 1h (including ~15 min heating up) in pure O$_2$ at a pressure $p_{O_2}$ = 15 Pa in the same vacuum chamber. After a spontaneous cooling down below 50 °C, a 50 nm thick SiO$_2$ film was deposited. The process, consisting of the V-W film deposition, its subsequent thermal oxidization and the SiO$_2$ film deposition was repeated four times. A thicker (160 nm), instead of 50 nm, top SiO$_2$ film was applied (see Fig. 1a) to enhance the antireflection (AR) effect, and to provide mechanical and environmental protection of the TC coating. The coatings were produced in an ultrahigh vacuum multimagnetron sputter device (ATC 2200-V AJA

International Inc.) equipped by unbalanced magnetrons with planar targets (diameter of 50 mm and thickness of 6 mm in all cases [5]). The V-W films were deposited by pulsed DC magnetron sputtering of a single V-W (4.0 wt.% corresponding to 1.14 at.% of W) target (99.95% purity) at the argon pressure $p_{Ar}$ = 1 Pa using a unipolar pulsed DC power supply (TruPlasma Highpulse 4002 TRUMPF Huettinger). The voltage pulse duration was 50 µs at a repetition frequency of 500 Hz and the deposition-averaged target power density (spatially averaged over the total target area) was 14.9 Wcm$^{-2}$. The SiO$_2$ films with $n$ of 1.46 and $k$ < 10$^{-3}$ at the wavelength of 550 nm were deposited by midfrequency bipolar dual magnetron sputtering of two Si (99.999% purity) targets at $p_{Ar}$ = 1 Pa and $p_{O_2}$ = 0.2 Pa using a bipolar dual power supply (TruPlasma Bipolar 4010 TRUMPF Huettinger). The voltage pulse duration was 10 µs at a repetition frequency of 50 kHz and the deposition-averaged target power density was approximately 8 Wcm$^{-2}$.

2.2. Coating characterization

The thickness of the as-deposited V-W and SiO$_2$ films and the optical constants of the SiO$_2$ films were measured by spectroscopic ellipsometry using the J.A. Woollam Co. Inc. VASE instrument. The room-temperature (25 °C) crystal structure of the multilayer coating was characterized by X-ray diffraction using a PANalytical X'Pert PRO diffractometer working with a CuKα (40 kV, 40 mA) radiation at a glancing incidence of 1°. The fine microstructure of the coating was investigated by transmission electron microscopy (TEM). Selected-area electron diffraction (SAED) patterns, and TEM and HRTEM images were recorded in a Hitachi H-9500 electron microscope that is attached with a Gatan SC-1000 Orius CCD camera (4008 × 2672 pixels) and an EDAX energy-dispersive spectroscopy (EDS) system for elemental analysis. The microscope was operated at 300 keV with a lattice resolution of 1 Å. Surface morphology of the individual nanocomposite layers was investigated by scanning electron microscopy (SEM) using a Hitachi SU-70 microscope. The coating transmittance ($T$) was measured by spectrophotometry using the Agilent CARY 7000 instrument with an in-house made heat/cool cell. The measurements were performed in the wavelength range of 300-2500 nm for $T_{ms}$ = -20 °C and $T_{mm}$ = 70 °C. Hysteresis curves were measured for $T$ at $\lambda$ = 2500 nm in the temperature range $T_m$ = -20 °C to 70 °C. The coating performance is quantified by means of integral luminous transmittance ($T_{lum}$), integral solar energy transmittance ($T_{sol}$) and their modulations ($\Delta T_{lum}$ and $\Delta T_{sol}$). The quantities are defined as

$$T_{lum}(T_m) = \frac{\int_{380}^{780} \varphi_{lum}(\lambda)\varphi_{sol}(\lambda)T(T_m,\lambda)d\lambda}{\int_{380}^{780} \varphi_{lum}(\lambda)\varphi_{sol}(\lambda)d\lambda},$$

$$\Delta T_{lum} = T_{lum}(T_{ms}) - T_{lum}(T_{mm}),$$

$$T_{sol}(T_m) = \frac{\int_{300}^{2500} \varphi_{sol}(\lambda)T(T_m,\lambda)d\lambda}{\int_{300}^{2500} \varphi_{sol}(\lambda)d\lambda},$$

$$\Delta T_{sol} = T_{sol}(T_{ms}) - T_{sol}(T_{mm}),$$

where $\varphi_{lum}$ is the luminous sensitivity of human eye and $\varphi_{sol}$ is the solar irradiance spectrum [12] at an air mass of 1.5. The average luminous transmittance is defined as $T_{lum} = [T_{lum}(T_{ms}) + T_{lum}(T_{mm})]/2$.

## 3. Results and discussion

### 3.1. Design and phase composition of the multilayer coating

Fig. 1 presents a cross-section TEM image and a SAED pattern taken from the high-performance TC coating with a thickness of ~520 nm on the SLG substrate. As can be seen in Fig. 1a, the coating is composed of a bottom $SiO_2$ layer with a thickness of ~100 nm, four layers of W-doped $VO_2$ nanoparticles (fabricated by annealing the as-deposited V-W films) dispersed in a $SiO_2$ matrix, which are separated by three $SiO_2$ layers with a thickness of ~50 nm, and a top $SiO_2$ layer with a thickness of ~160 nm. In Fig. 1b, the d-spacings of the diffraction spots 1 and 2 probably correspond to the (001) and (002) planes, respectively, of the monoclinic $\beta$-$Na_{0.33}V_2O_5$ phase (PDF # 01-084-8341 [13]), while that of the diffraction spots 3 (and 4) corresponds to the monoclinic $VO_2(M1)(011)$ planes (PDF # 04-003-2035) or to the tetragonal $VO_2(R)(110)$ planes (PDF # 01-073-2362), that of the diffraction spots 5 (and 6) corresponds to a superposition of the monoclinic $VO_2(M1)(\bar{2}02)$, $(\bar{2}11)$ and (200) planes or to the tetragonal $VO_2(R)(101)$ planes, and that of the diffraction spot 7 corresponds to a superposition of the monoclinic $VO_2(M1)(\bar{2}12)$ and (210) planes or to the tetragonal $VO_2(R)(111)$ planes.

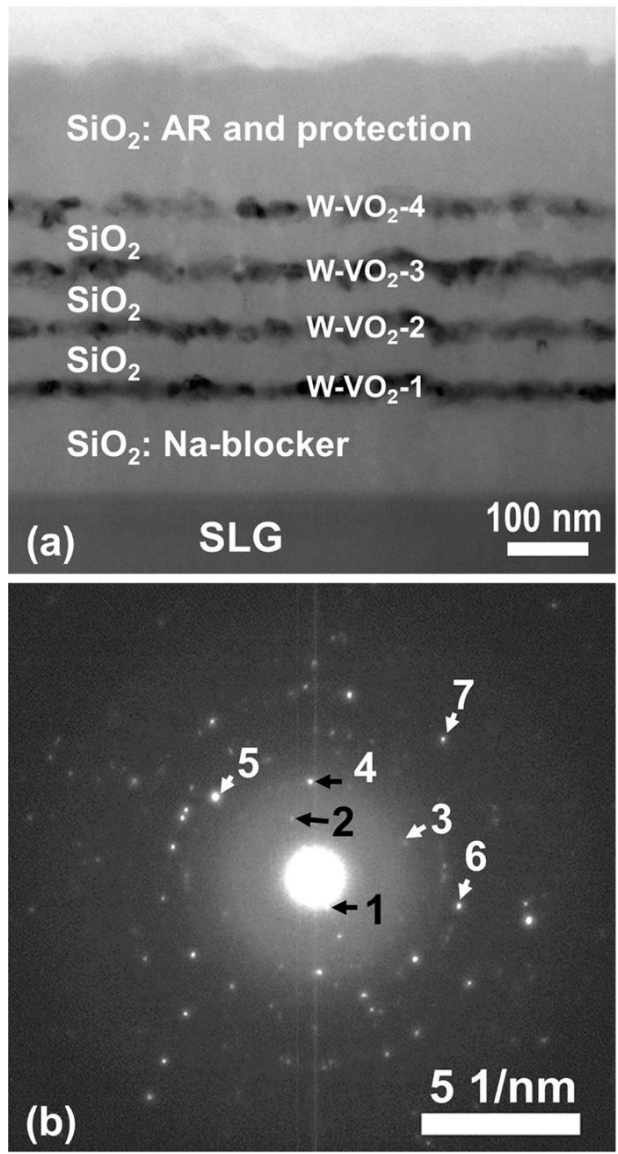

**Fig. 1**. (a) Cross-section TEM image of the multilayer coating with four layers of W-doped $VO_2$ nanoparticles dispersed in a $SiO_2$ matrix on 1 mm thick soda-lime glass. (b) SAED pattern taken from the coating with the d-spacings of the diffraction spots 1, 2, 3 (and 4), 5 (and 6) and 7 equal to 9.90 Å, 4.89 Å, 3.24 Å, 2.43 Å and 2.14 Å, respectively.

The crystalline phases identified by XRD in the TC coating can be seen in Fig. 2. They are in accordance with those identified in Fig. 1b. Besides a small contribution of the $\beta$-$Na_{0.33}V_2O_5$ phase caused by a combined effect of not fully blocked Na diffusion from the SLG substrate during not yet fully optimized thermal oxidization of the as-deposited V-W films, all other peaks are identified as diffraction peaks of either the low-temperature monoclinic $VO_2$(M1) phase or the high-temperature tetragonal $VO_2$(R) phase. These TC phases are difficult to distinguish, and they are actually expected to be present simultaneously [5] because the measurement temperature $T_m$ = 25 °C is close to the transition temperature $T_{tr}$ = 33 °C.

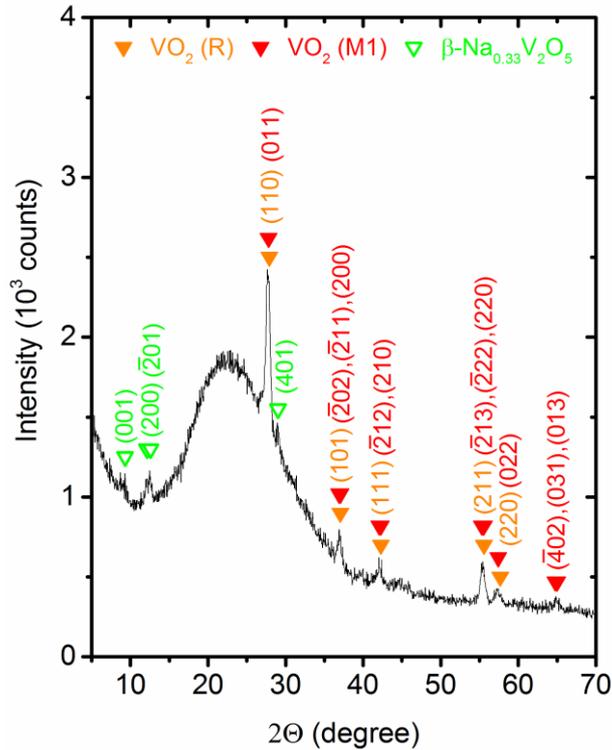

**Fig. 2**. X-ray diffraction pattern taken at $T_m$ = 25 °C from the multilayer coating with four layers of W-doped VO$_2$ nanoparticles dispersed in a SiO$_2$ matrix on 1 mm thick glass. The main diffraction peaks of VO$_2$(M1), VO$_2$(R) and β-Na$_{0.33}$V$_2$O$_5$ are marked.

3.2. Microstructure of single layers

Fig. 3 presents zoom-in cross-section TEM images of the W-VO$_2$-4 and W-VO$_2$-1 composite layers (see Fig. 1a), formed by two-dimensional arrays of the W-doped VO$_2$ nanoparticles in a SiO$_2$ matrix, with the corresponding EDS spectra and HRTEM images of single W-doped VO$_2$ nanoparticles. As can be seen in Figs. 3b,e, the X-ray EDS analysis confirmed that both layers exhibit presence of only V, W, Si, O and Na. The d-spacing of the lattice fringes related to the W-doped VO$_2$ nanoparticle in the W-VO$_2$-4 layer (Fig. 3c) is about 3.24 Å corresponding to the monoclinic VO$_2$(M1)(011) planes with $d$ = 3.21 Å or to the tetragonal VO$_2$(R)(110) planes with $d$ = 3.19 Å. The vertical size of this nanoparticle is ~42 nm. The d-spacing of the lattice fringes related to the W-doped VO$_2$ nanoparticle in the W-VO$_2$-1 layer (Fig. 3f) is about 2.14 Å corresponding to a superposition of the monoclinic VO$_2$(M1)($\bar{2}$12) and (210) planes with $d$ = 2.15 Å and 2.14 Å, respectively, or to the tetragonal VO$_2$(R)(111) planes with $d$ = 2.14 Å or even to the orthorhombic VO$_2$(P)(220) planes with $d$ = 2.18 Å (PDF # 00-025-1003). The vertical size of this nanoparticle is ~22 nm. Note that the vertical size of some of the nanoparticles is at least 4.2× larger than the V-W film thickness (42 nm compared to 10 nm), while the oxidation of V to VO$_2$ increases the volume per metal atom only 2.1× (from ~14 to ~29 Å [3]). This represents another fingerprint (in parallel to EDS) of the driving force toward dewetting. However, as can be seen in Figs. 3a,d, it is very probable that at least some of the W-doped VO$_2$ nanoparticles are connected to each other in the W-VO$_2$ layers [7].

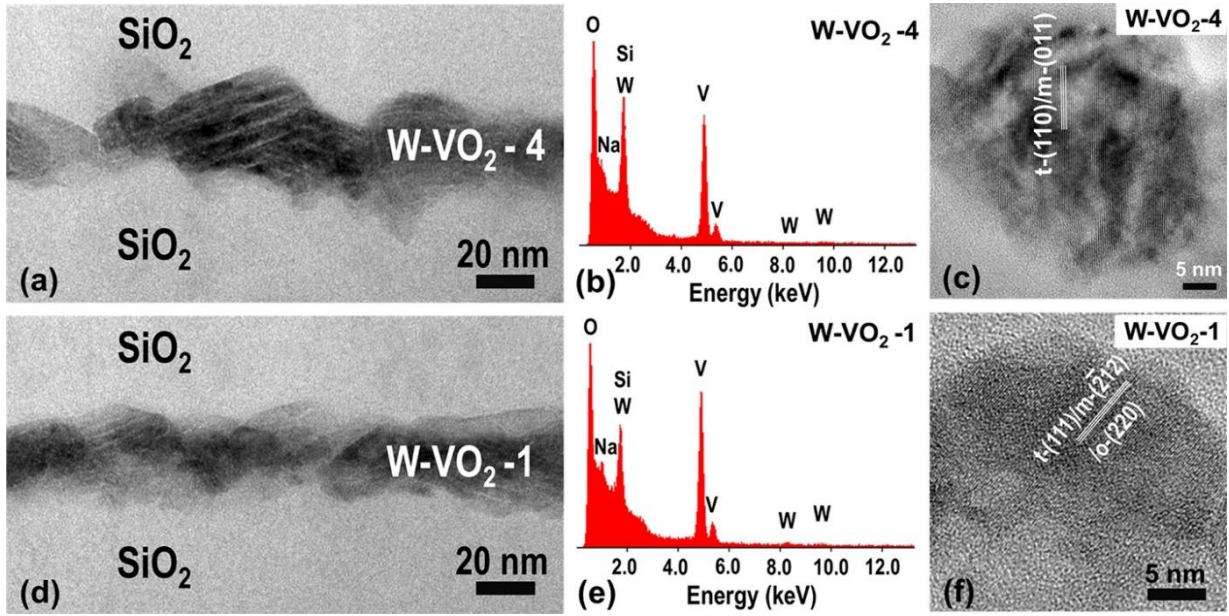

**Fig. 3.** (a) and (d) Zoom-in cross-section TEM images, and (b) and (e) EDS spectra of the W-$VO_2$-4 layer and the W-$VO_2$-1 layer, respectively. (c) and (f) HRTEM images of W-doped $VO_2$ nanoparticles from the same layers.

Fig. 4 presents top-view SEM images of the first layer of W-doped $VO_2$ nanoparticles fabricated on the Na-blocking $SiO_2$ layer. An image analysis showed that the diameter of most particles ranges from 41 nm to 55 nm, while the spacing between them is from 7 nm to 21 nm. Moreover, elongated particles with a length of up to 150 - 400 nm and even up to 1μm were identified. The surface morphology of this layer is different from the layers with well-separated, uniformly distributed $VO_2$ nanoparticles of the homogeneous size ≤ 200 nm, which were fabricated by annealing the as-deposited V films in a mixture of Ar and $O_2$ at $T_s \geq 500$ °C [8,10]. Recently, valuable results explaining the dewetting process in undoped $VO_2$ films and W-doped $VO_2$ films fabricated by annealing the as-deposited $VO_x$ and W-doped $VO_x$ films, respectively, in pure $O_2$ have been reported [14]. It has been shown that a higher temperature (≥ 600 °C) and a longer time of the annealing are required for dewetting of W-doped $VO_2$ nanoparticles.

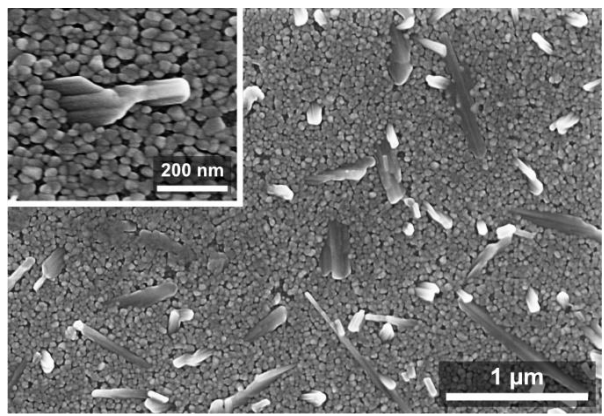

**Fig. 4.** SEM images (top view) of the first layer of W-doped $VO_2$ nanoparticles fabricated without any $SiO_2$ overlayer on $SiO_2$ layer blocking Na diffusion from the glass substrate (see Fig. 1a).

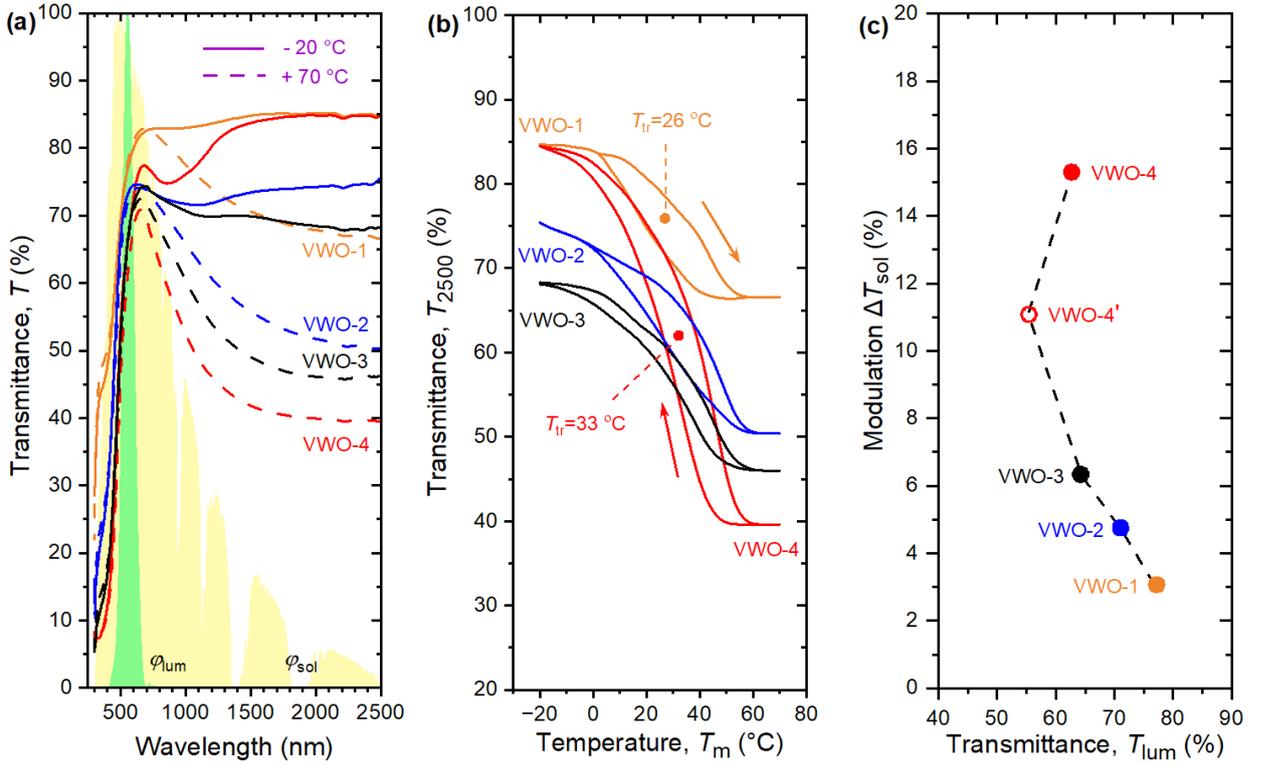

**Fig. 5.** (a) Spectral transmittance measured at $T_{ms} = -20$ °C and $T_{mm} = 70$ °C for four coatings with one (VWO-1), two (VWO-2), three (VWO-3) and four (VWO-4) layers of W-doped $VO_2$ nanoparticles with approximately 50 nm thick $SiO_2$ overlayers, except for the VWO-4 coating with approximately 160 nm thick $SiO_2$ overlayer, which were fabricated successively on $SiO_2$ layer blocking Na diffusion from the glass substrate (see Fig. 1a). The contours of the shaded areas represent the luminous sensitivity of the human eye ($\varphi_{lum}$) and the solar irradiance spectrum ($\varphi_{sol}$), normalized to maxima of 100%. (b) Temperature dependences of the transmittance at 2500 nm for the same coatings. The transition temperatures are given for the VWO-1 and VWO-4 coatings. (c) Average integral luminous transmittance and the modulation of the solar energy transmittance for the same coatings and for the VWO-4´ coating with approximately 50 nm thick $SiO_2$ overlayer.

### 3.3. Thermochromic properties

The optical performance resulting from the presented coating architecture and preparation protocol (keeping in mind the exceptionally industry-friendly low preparation temperature) is summarized in Fig. 5 and Table 1. Indeed, our methodology allowed us to prepare strongly TC coatings (large modulation in the infrared in Fig. 5a) with lowered transition temperature ($T_{tr}$ of at most 33 °C in Fig. 5b). Although the present setup (single alloy target with 1.14 at.% of W) did not allow fine tuning of $T_{tr}$, the $T_{tr}$ gradient of at least -24 °C per at.% of W in the target (assuming $T_{tr}$ ~60 °C of thin-film $VO_2$ [3], while $T_{tr} = 68$ °C of bulk $VO_2$ would lead to an even larger gradient) is larger than most gradients reported previously [3]. This indicates a successful incorporation of W into the metal sublattice of $VO_2$, instead of a significant segregation at the nanoparticle surface (which has been reported [14] in parallel to a somewhat smaller $T_{tr}$ gradient).

Quantitatively speaking, the gradual increase of the number of TC layers from 1 to 4 leads to monotonically but relatively slowly decreasing $T_{lum}$, from 76.3-77.1% through 70.5-71.9% and 63.9-64.9% to 53.8-56.8%. While these values are already affected by the lower $n$ and in turn lower reflectance of $SiO_2$-containing composite TC layers compared to compact homogeneous W-doped $VO_2$, altering the AR effect by the transition from 50 nm to 160 nm $SiO_2$ overlayer leads to a further enhancement of $T_{lum}$ from 53.8-56.8% to 60.1-65.4%. This has to be interpreted in the context given by total thickness of the four V-W layers of ~40 nm, corresponding to ~84 nm of W-doped $VO_2$ with the same areal density of metal atoms. The present $T_{lum}$ = 60.1-65.4% of the coating based on doped $VO_2$ nanoparticles is actually higher than $T_{lum}$ of coatings [3,5] based on compact pure or doped $VO_2$ layers having thicknesses well below 84 nm, and dramatically higher than $T_{lum}$ predicted [3,5] for coatings based on compact layers having a similar thickness of 80 nm. Moreover, there is a potential for further $T_{lum}$ enhancement by fine tuning of $n$ and thickness of the AR overlayer, while the aforementioned predictions assumed an optimized one. One reason for this success, which increases the potential to achieve high $\Delta T_{sol}$ at sufficient $T_{lum}$, is the non-linear dependence of $k$ of the composite TC layer on its composition. To give an example, assuming $SiO_2$ matrix with spherical inclusions of W-doped $VO_2$ (this is not to claim such a perfect dewetting in the present case), a calculation based on the Maxwell-Garnett effective medium approximation and equal volume fractions of both these phases yields $k$ at $\lambda$ = 550 nm of only ~63% of the corresponding arithmetic mean.

In parallel to the evolution of $T_{lum}$, the gradual increase of the number of TC layers from 1 to 4 leads to increasing $\Delta T_{sol}$ from 3.1% through 4.7% and 6.3% to 11.1%, and the transition from 50 nm to 160 nm $SiO_2$ overlayer leads to its further enhancement to 15.3%. While the basic factor is the increasing transmittance modulation in the infrared with increasing amount of the TC phase (again, this amount is allowed to be large because of the aforementioned preservation of high $T_{lum}$), there are three more phenomena to note. First, there is almost monotonically increasing transmittance modulation also in the visible, from $\Delta T_{lum}$ = -0.8% to 5.3% (seemingly low but multiplied by high $\varphi_{sol}$ and in turn responsible for about a quarter of the $\Delta T_{sol}$ enhancement). Second, Fig. 5a shows that the 160 nm thick $SiO_2$ overlayer leads to exceptionally non-monotonic low-temperature $T(\lambda)$ with a (second-order) maximum in the visible complemented by a (first-order) maximum in the infrared, increasing the contribution of infrared wavelengths to $\Delta T_{sol}$. Third, the absorption in the high-temperature state, and in turn $\Delta T_{sol}$, is arguably enhanced by LSPR at the boundary between metallic W-doped $VO_2$(R) nanoparticles and dielectric $SiO_2$. The resonance condition is prone to be close to fulfilled in the near infrared [2,3]. This explains that the coating performance at a given areal density of metal atoms is better than that of coatings based on compact pure or doped $VO_2$ layers (including coatings which utilize the previous two phenomena [3,5]) not only in terms of $T_{lum}$ (previous paragraph) but also in terms of $\Delta T_{sol}$. A case can be made that the surface plasmon resonance is not completely localized due to the interconnection of some nanoparticles (Figs. 3a,d) and that the potential of this phenomenon is even larger. Such non-idealities may be behind different high-temperature $T(\lambda)$ of different $VO_2$-based coatings utilizing LSPR, sometimes [8,10] exhibiting a minimum in the near infrared, sometimes [9] (consistently with our result) monotonically decreasing in the near infrared. Collectively, the presented coating characteristics show that while the well known [2,3] tradeoff between $T_{lum}$ and $\Delta T_{sol}$ due to varied amount of the TC material takes place also in the present case, the specific pairs of these values (Fig. 5c) are beyond anything reported until now at such a low $T_{tr}$.

**Table 1**

Integral luminous and solar energy transmittance [$T_{lum}(T_m)$ and $T_{sol}(T_m)$, respectively] measured at $T_{ms} = -20$ °C and $T_{mm} = 70$ °C, together with the corresponding modulations $\Delta T_{lum}$ and $\Delta T_{sol}$, and the transition temperature, $T_{tr}$, corresponding to the middle of hysteresis curves (Fig. 5b), for five coatings with one (VWO-1), two (VWO-2), three (VWO-3) and four (VWO-4′ and VWO-4) layers of W-doped $VO_2$ nanoparticles with approximately 50 nm thick $SiO_2$ overlayers, except for the VWO-4 coating with approximately 160 nm thick $SiO_2$ overlayer, which were fabricated successively on $SiO_2$ layer blocking Na diffusion from the soda-lime glass substrate (Fig. 1a).

| Coating | $T_{lum}(T_{ms})$ (%) | $T_{lum}(T_{mm})$ (%) | $\Delta T_{lum}$ (%) | $T_{sol}(T_{ms})$ (%) | $T_{sol}(T_{mm})$ (%) | $\Delta T_{sol}$ (%) | $T_{tr}$ |
|---|---|---|---|---|---|---|---|
| VWO-1 | 76.3 | 77.1 | -0.8 | 76.8 | 73.7 | 3.1 | 26 |
| VWO-2 | 71.9 | 70.5 | 1.4 | 67.2 | 62.5 | 4.7 | 31 |
| VWO-3 | 64.9 | 63.9 | 1.0 | 62.4 | 56.1 | 6.3 | 32 |
| VWO-4′ | 56.8 | 53.8 | 3.0 | 55.8 | 44.7 | 11.1 | 33 |
| VWO-4 | 65.4 | 60.1 | 5.3 | 66.1 | 50.8 | 15.3 | 33 |

## 4. Conclusions

We present a high-performance (the average $T_{lum} = 62.8\%$ and $\Delta T_{sol} = 15.3\%$) TC $VO_2$-based coating with a decreased $T_{tr} = 33$ °C and enhanced environmental protection, which was prepared on conventional glass at a low $T_s = 350$ °C without opening the vacuum chamber to atmosphere. It is formed by four layers of W-doped $VO_2$ nanoparticles dispersed in $SiO_2$ matrix. Such a combination of properties, together with the low temperature during a three-step preparation process (magnetron sputter depositions and postannealing in oxygen), fulfill the requirements for large-scale implementation on building glass and have not been reported yet. Necessary reduction in the number of layers and shortening the preparation process would move us closer to reducing the energy consumption of buildings by applying this kind of coatings on windows and glass facades.

**Data availability**

Data will be made available on request.

**Acknowledgments**

This work was supported by the project Quantum materials for applications in sustainable technologies (QM4ST), funded as Project no. CZ.02.01.01/00/22_008/0004572 by Programme Johannes Amos Commenius, call Excellent Research, and the National Science Foundation Partnership for Research and Education in Functional Materials (NSF DMR-2425164).